# Scanning Tunneling Microscopy of an Air Sensitive Dichalcogenide Through an Encapsulating Layer

*Jose Martinez-Castro[1,†], Diego Mauro[1,2,†], Árpád Pásztor[1,†], Ignacio Gutiérrez-Lezama[1,2], Alessandro Scarfato[1], Alberto F. Morpurgo[1,2], Christoph Renner[1,]\**

[1]Department of Quantum Matter Physics, University of Geneva, 24 Quai Ernest-Ansermet, CH-1211 Geneva 4, Switzerland.

[2]Group of Applied Physics (GAP), University of Geneva, 24 Quai Ernest-Ansermet, CH-1211 Geneva 4, Switzerland.

ABSTRACT:

Many atomically thin exfoliated 2D materials degrade when exposed to ambient conditions. They can be protected and investigated by means of transport and optical measurements if they are encapsulated between chemically inert single layers in the controlled atmosphere of a glove box. Here, we demonstrate that the same encapsulation procedure is also compatible with scanning tunneling microscopy (STM) and spectroscopy (STS). To this end, we report a systematic STM/STS investigation of a model system consisting of an exfoliated 2H-NbSe$_2$ crystal capped with a protective 2H-MoS$_2$ monolayer. We observe different electronic coupling between MoS$_2$ and NbSe$_2$, from a strong coupling when their lattices are aligned within a few degrees to essentially no coupling for 30° misaligned layers. We show that STM always probes intrinsic NbSe$_2$ properties such as the superconducting gap and charge density wave at low temperature when setting the tunneling bias inside the MoS$_2$ band gap, irrespective of the relative angle between the NbSe$_2$ and MoS$_2$ lattices. This study demonstrates that encapsulation is fully compatible with STM/STS investigations of 2D materials.

Exfoliation of layered van der Waals (vdW) materials has proven to be a remarkably simple technique to produce high-quality crystals of many different compounds that are only one or a few atoms thick.[1–8] These atomically thin crystals –or 2D materials– possess new interesting properties that can be very different from those of their parent bulk compounds and can depend very sensitively on the precise number of atomic layers.[4,9] As such, 2D materials disclose a vast platform for the investigation of new physical phenomena that were not accessible to experiment until now. Examples include Dirac fermions in monolayer (ML) graphene[1,10], gate-tuning of the band structure of a 2D material first shown in bilayer graphene[11], phenomena originating from the Berry curvature in the band structure of semiconducting ML transition metal dichalcogenides (TMDs)[12–14], and magnetism and superconductivity in the truly 2D limit.[7,15,16]

Most experimental studies reported so far have been performed on materials that are chemically stable in air because this drastically simplifies their manipulation and device fabrication for a broad variety of experimental techniques. However, many 2D materials tend to degrade when exposed to air, and considerable efforts are deployed to protect exfoliated crystals and enable their characterization and use under ambient conditions. An effective strategy is to exfoliate and manipulate the atomic layers in a glove box and then encapsulate them with another inert single layer crystal[6–8,17,18], for example graphene, MoS$_2$ or hBN. Even though the procedure is complex, encapsulation is remarkably efficient, enabling air-sensitive 2D materials to be safely exposed to air. Encapsulation has been key to a number of remarkable experiments, for example the observation of a 2D topological insulating state in ML WTe$_2$[6], the investigation of superconductivity in ML 2H-NbSe$_2$[19], and the observation of 2D ferromagnetism in ML Cr$_2$Ge$_2$Te$_6$ and CrI$_3$.[7,8]

The controlled heterostructure assembly and encapsulation of vdW MLs represents an impressive technical achievement and demonstrates an unprecedented level of control of matter at the atomic scale. While these techniques have been successfully used to prepare samples and devices for transport and optical measurements, their compatibility with surface probes like scanning tunneling microscopy (STM) and angle-resolved photoemission spectroscopy (ARPES) is yet to be verified. Both are primarily sensitive to the outermost layers of the system under investigation, and covering a 2D material with an encapsulating layer may potentially impede their use altogether. In addition, the encapsulating layer may affect the electronic properties of the underlying 2D material, due to their mutual interaction.[20–24] If so, even if STM and ARPES measurements were technically possible, it would be necessary to understand to what extent the measurements are representative of the properties of the encapsulated 2D material.

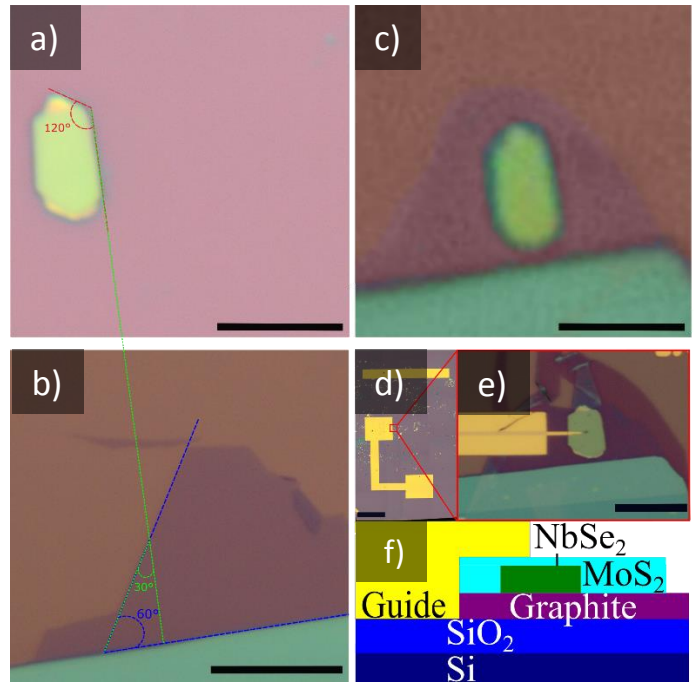

**Figure 1. Optical images and schematic description of the devices for STM measurements.** Optical microscope image of an exfoliated 2H-NbSe$_2$ crystal **a)** and of an exfoliated 2H-MoS$_2$ crystal composed of a ML and a bulk part **b)**. **c)** Optical microscope image of the 2H-NbSe$_2$ crystal shown on a) encapsulated under the ML part of the 2H-MoS$_2$ shown in b). **d)** Optical microscope image of the entire device fabricated on a Si/SiO$_2$ substrate. The device is composed of a strip (upper side) to align the sample with the XY piezoelectric drives, a tip landing pad (center square) and a connecting pad (lower square). **e)** Closeup of the contacted heterostructure showing the contact and the tip guide. Scale bar: 5 μm, except panel d) 1 mm. **f)** Schematic cross section of the heterostructure device itself.

To assess the possibility to perform STM imaging and spectroscopy of encapsulated 2D materials, we investigate a 90 nm thick 2H-NbSe$_2$ crystal (Fig. 1a), whose properties match those of bulk 2H-NbSe$_2$ (see Supporting Information 1 (SI 1) for more details), capped with an exfoliated 2H-MoS$_2$ ML (Fig. 1b). This choice is motivated by the excellent knowledge that we have of both materials. NbSe$_2$ is a prototypical layered superconductor with a nearly commensurate *ab*-plane ≈3a×3a CDW modulation developing below $T_{CDW}$ = 32 K and superconductivity developing below $T_C$ = 7.2 K.[25,26] These two macroscopic quantum phases make NbSe$_2$ ideal to evaluate atomic resolution imaging and spectroscopy through a capping layer. MoS$_2$ was chosen as capping layer because of its chemical stability, the availability of large, easy to detect and manipulate exfoliated MLs and its sufficiently large band gap at the Fermi level. Stability and size are obvious criteria for an effective capping layer. The band gap is required to allow STM to access low energy spectral features of NbSe$_2$. Indeed, in elastic tunneling through a potential barrier, electrons tunnel from filled states in the tip to empty states in the sample at positive sample bias, and vice versa for negative sample bias. When regulating the tip position above MoS$_2$ at a bias voltage $V_{set}$ (defined as set point) outside the band gap, the sample states are provided primarily by the MoS$_2$ capping layer and we expect to probe predominantly MoS$_2$ derived properties. By selecting $V_{set}$ inside the MoS$_2$ band gap, however, no states are available in the capping layer at low temperature and tunneling must in principle occur between the tip and NbSe$_2$. This allows accessing the low energy properties (within the band gap of MoS$_2$) of NbSe$_2$.

NbSe$_2$/MoS$_2$ heterostructures were assembled in a glove box by dry pick-up and transfer[18] of 2H-MoS$_2$, 2H-NbSe$_2$ and few-layer-Graphene (FLG) exfoliated onto different SiO$_2$/Si substrates. To explore the impact of the crystalline alignment, MoS$_2$ and NbSe$_2$ were stacked with two different angles α between their basal plane lattice vectors: either α=30° corresponding to the maximum possible misalignment (Fig. 1c) or α=3°, close to perfect alignment. These structures are placed on a FLG back electrode and contacted by means of conventional nanofabrication techniques (electron beam lithography, metal evaporation and lift-off). Specifically, we first deposit large enough gold pads to both contact the device and land the probe tip safely near the heterostructure. We also deposit a gold reference bar to align the device with the XY scanning directions of the STM using the optical setup of the UHV STM chamber (Fig. 1d). In a second step, we deposit a 1 μm wide and 40 nm high gold strip to guide the tip along the contact from the landing pad to the device. An optical image and a schematic drawing of the device structure are displayed in Figs. 1e and 1f, respectively. Further details of the device fabrication are given in the methods section.

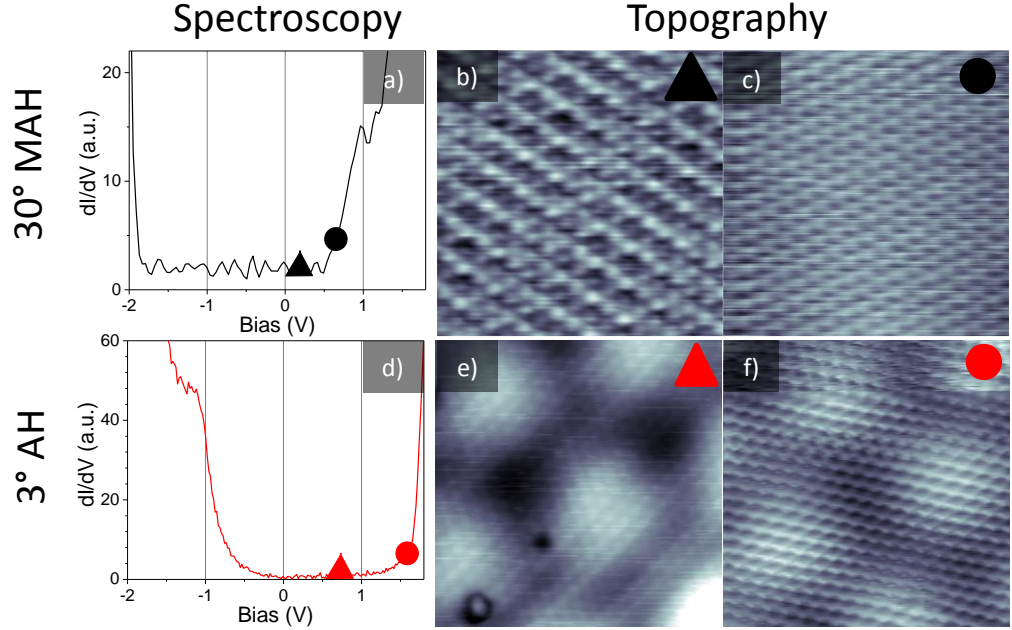

**Figure 2. Tunneling spectra and topographic STM images of the MoS$_2$/NbSe$_2$ heterostructures. a)** Tunneling spectrum of a 30° MAH device at T=77 K ($V_{set}$=2.5 V, $I_{set}$=200 pA). The black triangle and circle mark the energies of the topographic images in panels b) and c) respectively. **b)** Topographic STM image of the 30° MAH acquired inside the MoS$_2$ bandgap ($I_{set}$=10 pA, $V_{set}$=10 mV). **c)** Topographic STM image of the 30° MAH acquired outside the MoS$_2$ bandgap ($I_{set}$=2 nA, $V_{set}$=600 mV). **d)** Tunneling spectrum of a 3° AH taken at T=1.14 K ($V_{set}$=-1.5 V, $I_{set}$=200 pA mV). The red triangle and circle mark the energies of the topographic images in panels e) and f) respectively. **e)** Topographic STM image of the 3° AH acquired inside the MoS$_2$ bandgap ($I_{set}$=10 pA, $V_{set}$=800 mV). **f)** Topographic STM image of the 3° AH acquired outside the MoS$_2$ bandgap ($I_{set}$=10 pA, $V_{set}$=1.6 V). Size of all STM images is 5 nm × 5 nm.

The heterostructures were characterized by STM at low and high $V_{set}$. The relative alignment between 2H-MoS$_2$ and 2H-NbSe$_2$ layers inferred from the optical microscope images was further confirmed by measuring the periodicity of the resulting moiré pattern (See SI 2 for more details). To quantify the impact of the layer alignment on the electronic properties of a given heterostructure, we first measure the MoS$_2$ bandgap for the two different values of α at high $V_{set}$. Differential tunneling conductance spectra measured on the 30° misaligned heterostructure (MAH) show a band gap of ~2.3 eV and electron doping character (i.e., the Fermi level is aligned close to the bottom of the conduction band Fig. 2a), in agreement with gap values previously reported for MoS$_2$ monolayer.[27,28] The band gap measured for the same setpoint on the 3° aligned heterostructure (AH) has a similar amplitude but is shifted by nearly 1 eV toward the valence band (Fig. 2d). This band shift provides a first indication of significant electronic coupling between the capping layer and NbSe$_2$ in the 3° AH, which is absent in the 30° MAH.

To further characterize the electronic coupling for the two different crystal alignments, we perform STM imaging of both devices at $V_{set}$ inside and outside the MoS$_2$ ML bandgap. Imaging the 30° MAH at $V_{set}$ inside the MoS$_2$ gap, we observe atomic periodicity and a Moiré pattern (Fig. 2b) due to contributions from both MoS$_2$ and NbSe$_2$ lattices. Increasing the bias voltage $V_{set}$ outside the MoS$_2$ gap, the observed pattern becomes simpler, the STM only resolves a simple triangular atomic lattice corresponding to MoS$_2$ with a=3.13 Å (Fig. 2c). Imaging the 3° AH yields a strikingly different result. In this case, the Moiré pattern is always observed independently of $V_{set}$ (Fig. 2e and 2f) , whether $V_{set}$ is inside or outside the capping layer gap. We observe the same effects at both positive and at negative sample biases.

The tunneling spectroscopy and topographic imaging discussed above both lead to the conclusion that the measurements strongly depend on the misalignment angle α. At high $V_{set}$, we only sense MoS$_2$ states on the 30° MAH while we get a complex response with contributions from both MoS$_2$ and NbSe$_2$ states on the 3° AH. The latter observation provides an additional indication of a strong coupling of the MoS$_2$ ML to NbSe$_2$ when the two crystals are aligned, so much so that it is impossible to get an image of the sole MoS$_2$ capping layer, even at bias voltages outside the MoS$_2$ gap. In the following, we show that despite these different coupling regimes, the contributions of the electronic properties of NbSe$_2$ can be extracted from the experiments irrespective of the misalignment angle with MoS$_2$, which at low $V_{set}$ acts as a spatially modulated barrier that the electrons have to tunnel through.

For a quantitative analysis and understanding of the STM micrographs, we start by discussing simulations of real space STM images whose fourier transorm (FT) is reproducing the FT of the experimental micrographs. At low bias, with $V_{set}$ within the MoS$_2$ gap, STM images systematically resolve a superstructure for both heterostructures. To understand this observation, we propose that MoS$_2$ and NbSe$_2$ both contribute to the local tunneling current $I(x,y,z)$ calculated as $I(x,y,z) \approx V_{set}\rho_s(x,y)e^{-2\kappa(x,y)\cdot z}$. Here, $\rho_s(x,y)$ is the local density of states of the NbSe$_2$ surface and $\kappa(x,y)$ quantifies the local tunneling barrier height, which we assume to be spatially modulated by the capping layer atomic structure (Fig. 3a). $\rho_s(x,y)$ and $\kappa(x,y)$ are modelled as the sum of harmonic functions with the periodicity of the respective materials, respecting the three-fold crystal symmetry. We define the *x* and *y* coordinates parallel to the *ab*-basal plane and the *z*-axis perpendicular to it (see SI 3 for more details of the modeled images and for a representative plot of the function used to model the spatial modulation).

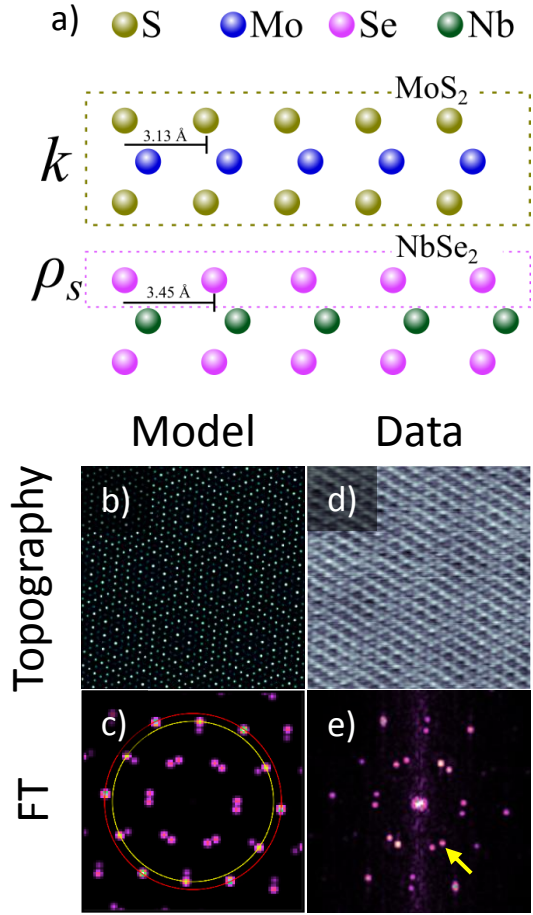

**Figure 3. Low bias STM imaging of the 30° misaligned heterostructure at 77 K. a)** Cross-section model of the $MoS_2$ / $NbSe_2$ heterostructure to illustrate the contributions to the simulated current image. The $MoS_2$ lattice defines $\kappa$ while $\rho_s$ is determined by the top Se layer of $NbSe_2$. **b)** Simulated real space current map and **c)** corresponding FT. Red and yellow circles correspond to the lengths of the 2H-$MoS_2$ and 2H-$NbSe_2$ lattice q-vectors, respectively. **d)** 10 nm × 10 nm topographic STM image ($I_{set}$=10 pA, $V_{set}$=10 mV) and **e)** corresponding FT. Yellow arrow marks the superstructure peaks.

The FT of the simulated $I(x,y,z)$ map of a 30° MAH (Fig. 3c) is reproducing very well the main features of the experimental FT in Fig. 3e when we assume that $\kappa(x,y)$ is modulated by a simple triangular lattice (i.e. considering only the Bravais-lattice of $MoS_2$). The FT peaks located on the red and yellow circles in Fig.3c correspond to the reciprocal lattice vectors of $MoS_2$ and $NbSe_2$, respectively. Superstructure peaks originating from the Moiré pattern (highlighted by a yellow arrow), and other fine details are also well reproduced by this model.

So far, we have analyzed the STM micrographs in reciprocal space, finding a very satisfactory correspondence between the experiments and the model FT. We now turn to real space STM images, which allow us to refine the model of STM imaging through a ML capping layer. The tunnel junction between the STM tip and the surface has atomic scale dimensions. This means that the tunneling probability depends on whether the tip is positioned over a Mo or a S atom. To take this into account, we consider the complete three-layer structure of the $MoS_2$ unit cell (i.e. Bravais-lattice and base) and modulate $\kappa(x,y)$ by two triangular lattices, one representing the topmost Mo atoms and the other the vertically aligned bottom and top S atoms in the capping layer (Fig. 3a). The simulated constant current STM image (Fig. 3b) indeed reproduces the observed Moiré pattern and the circular footprint of the atoms in the experimental image (Fig. 3d). In contrast, if we only consider the Bravais lattice of $MoS_2$ and modulate $\kappa(x,y)$ with a single triangular lattice, the atoms in the simulated image have an elongated shape not observed in the data (see SI 4). Note that the FT is essentially independent on these two choices of $\kappa(x,y)$. This analysis shows that the full structure of the capping layer must be considered to understand the details of tunneling into the protected 2D material.

Now that we understand the STM topographic contrast of the heterostructures considered here, we turn to the final question, namely whether STM can access intrinsic electronic properties of the capped 2H-$NbSe_2$ crystal, beyond the simple imaging of the $NbSe_2$ lattice demonstrated above. To this end, we analyze constant height STM images of the 3° AH (Fig.4a) and its FT (Fig. 4b) measured at $V_{set}$=10 mV and at 1.14 K, well below the CDW phase transition. We focus on constant height images, because our model is a simulation of the position dependent total tunneling current. Our simulated current map (Fig.4c) and the corresponding FT (Fig.4d) reproduce all the features of the experimental current map and corresponding FT, provided that we include an additional 3a×3a periodic modulation of the density of states of $NbSe_2$ $\rho_s(x,y)$, which is a manifestation of the CDW modulation present in the $NbSe_2$ lattice.[25] If we ignore this CDW modulation, the FT (Fig. 4f) extracted from the simulated current map (Fig. 4e) does not reproduce the CDW peaks present in the experimental FT (Fig. 4b). To further confirm the detection of a CDW modulation independent of the twist angle, we prepared a heterostructure with an intermediate misalignment angle α = 12°. Accordingly, CDW peaks are observed in the corresponding FT and confirmed in the simulated $I(x,y,z)$ FT (see SI 5).

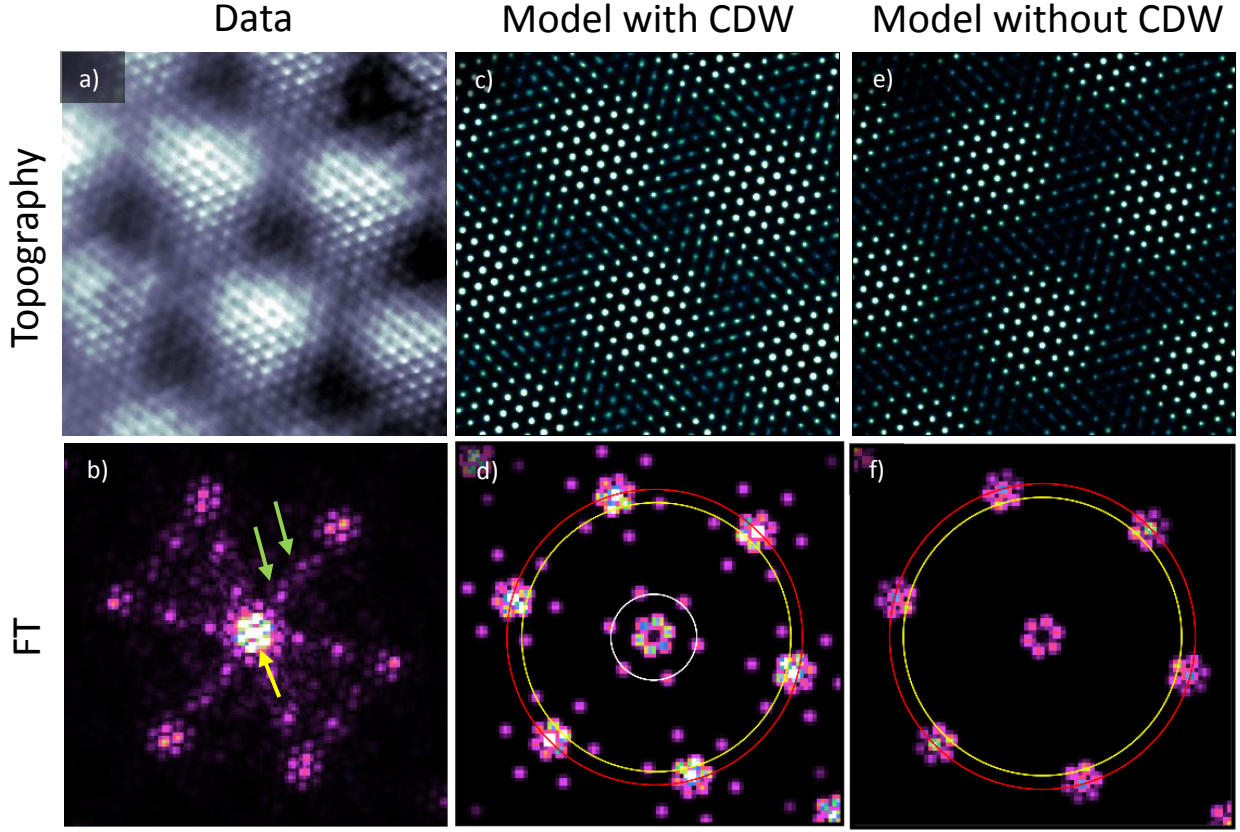

**Figure 4. Low bias STM imaging of the 3° aligned heterostructure at 1.14 K. a)** 5 nm × 5 nm constant height STM topography ($I_{set}$=10 pA, $V_{set}$=10 mV) and **b)** corresponding FT. Green and yellow arrows highlight CDW and Moiré pattern points, respectively. **c)** Simulated real space current map adding the 3x3 CDW modulation in NbSe$_2$ and MoS$_2$ lattice contributions and **d)** corresponding FT. The red and yellow circles mark the amplitudes of the MoS$_2$ and NbSe$_2$ lattice q-vector, respectively. The white circle indicates the amplitude of the 3x3 CDW q-vector. **e)** Simulated real space current map taking only the NbSe$_2$ and MoS$_2$ lattices into account, and **f)** corresponding FT.

The above example demonstrates our ability to extract precise topographic information about the NbSe$_2$ CDW through the MoS$_2$ capping layer. In NbSe$_2$, the CDW is susceptible to a reversible 3Q to 1Q transition driven by local strain.[29,30] The vast majority of our STM micrographs display the standard 3Q pattern shown in Fig.4, indicating there is no strain in our heterostructures. However, we do occasionally observe small regions with a 1Q CDW. One example is shown in Fig. 5a for a 3° AH, with two anisotropic peaks resolved in the FFT (Fig. 5b, arrow). Filtering the MoS$_2$ and superstructure contributions enables us to assign the anisotropic peaks to a strong 1D CDW component (Fig. 5c). The modulation period of this 1Q phase is slightly longer than that of the 3Q phase, in perfect agreement with previous findings on bare NbSe$_2$ surfaces[29,30] (Fig.5d). The 1Q phase develops in a limited region of the imaged surface while the rest is supporting the usual 3Q CDW. This result is remarkable, showing that all the features observed on bare NbSe$_2$ are also observed in presence of the capping layer.

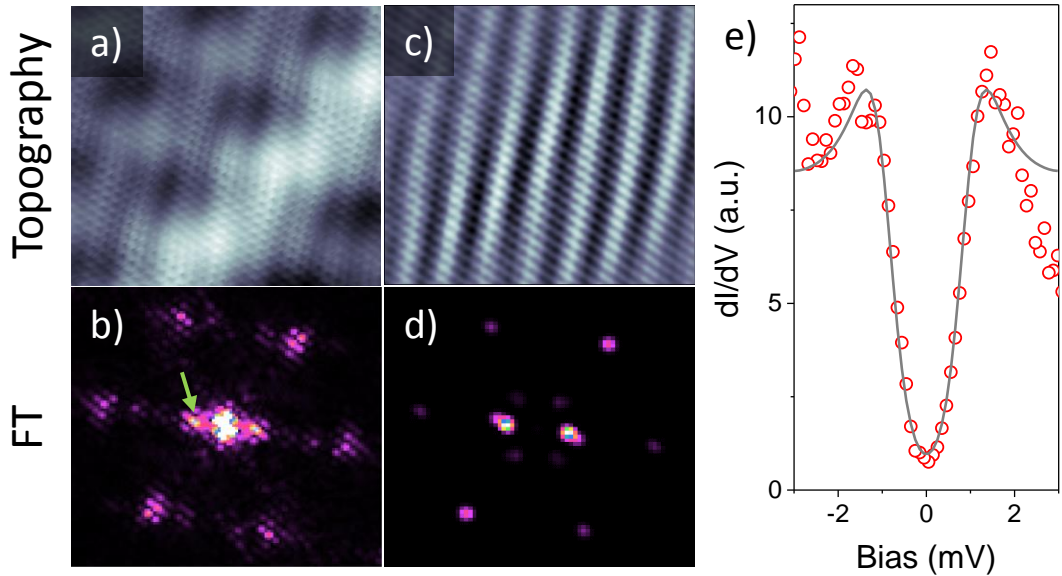

**Figure 5. Low bias STM imaging of a 1Q periodic charge modulation and NbSe$_2$ superconducting gap spectroscopy of a 3° aligned heterostructure. a)** 7.2 nm × 7.2 nm topographic STM image at 1.14 K ($V_{set}$=100 mV, $I_{set}$=10 pA) and **b)** corresponding FT. The green arrow points at the 1Q periodic modulation. **c)** Filtered topography by selecting only the atomic lattice, the 3Q and 1Q FT components of NbSe$_2$ and **d)** corresponding FT. **e)** The empty circles are the measured low energy tunneling differential conductance (T=1.14 K) showing the superconducting gap of 2H-NbSe$_2$. The solid line is a BCS fit with the superconducting gap Δ=0.98 meV and an effective temperature $T_{eff}$=2.52 K ($V_{set}$=5 mV, $I_{set}$=500 pA, $V_m$=100 µV).

Having established the possibility to probe NbSe$_2$ specific topographic features through the MoS$_2$ capping ML, including the CDW, it remains to be shown that tunneling spectroscopy of the protected NbSe$_2$ crystal is also possible. To address this point, we perform tunneling spectroscopy on the 3° AH and 12° MAH at 1.14 K, well below its superconducting transition temperature. Regulating the tip at $V_{set}$=5 mV, inside the MoS$_2$ band gap, we measure a gap at the Fermi level[27,28] (Fig. 5e and SI 5). The gap amplitude and line shape are consistent with the BCS expectations for superconducting NbSe$_2$. The only discrepancy is an effective temperature of 2.5 K instead of the 1.14 K indicated by the thermometer, most likely due to limited filtering of the electrical connections (the higher effective temperature is definitely not due to the MoS$_2$ capping layer, since we observe the same difference on bare NbSe$_2$).

In summary, we have demonstrated the possibility of studying the electronic properties of 2H-NbSe$_2$ by STM through an encapsulating 2H-MoS$_2$ monolayer. This key result paves the way for scanning probe microscopy studies of artificial heterostructures. From a physics point of view, our results show that the MoS$_2$ capping monolayer adds a non-trivial contribution to the vacuum tunneling barrier between tip and sample, significantly modulating the tunneling probability on atomic length scale. This observation may have implications for STM studies of other systems where tunneling occurs through a complex non-metallic atomic structure, for example tunneling into the CuO$_2$ layer in Bi$_2$Sr$_2$CaCu$_2$O$_8$.

**Methods**. MoS$_2$, few-layer-Graphene (FLG) and NbSe$_2$ are exfoliated onto different SiO$_2$/Si substrates (oxyde thickness=285 nm) and layers of the desired thickness are identified by inspection made with an optical microscope. A glove box is used for NbSe$_2$ exfoliation and later assembly of the heterostructure. The heterostructure is assembled using the so-called dry pick-up transfer method.[18] An MoS$_2$ ML is picked up with the help of a polycarbonate (PC) film. Next, with the aid of an optical microscope and a micromanipulator, we place the MoS$_2$ ML on top of the NbSe$_2$ with the desired misalignment angle. Finally, the MoS$_2$ ML/NbSe$_2$ stack is picked up and placed on top of a FLG single crystal. The complete heterostructure is then submerged in chloroform to remove the PC film used during the transfer.

Conventional electron-beam lithography (EBL), electron-beam metal evaporation and lift-off techniques are used to define the contacts, landing pads and tip guides. The former consists of a 10/70 nm thin film Ti/Au while the guide is made of a 10/30nm Ti/Au film.

During the lithographic processes, the heterostructure is continuously exposed to chemical impurities coming mostly from the poly-methyl methacrylate (PMMA) resist used to define the contacts. PMMA residues are removed from the surface in a controlled way by an atomic force microscope (AFM) "ironing" process.[31] It consists in scanning the heterostructure with an AFM tip in contact mode to displace surface impurities to the edges of the heterostructure.

Scanning tunneling experiments were carried out in a low temperature *Specs Tyto-STM* with a base pressure better than $2 \cdot 10^{-10}$ mbar with tips electrochemically etched from an annealed tungsten wire. The bias voltage was applied to the sample. Tunneling I(V) and differential conductance dI/dV(V) spectra were acquired simultaneously using a standard lock-in technique. dI/dV tunneling spectra were acquired with a bias modulation of 7 mV or 70 µV rms at 854.7 Hz to measure the semiconducting gap or the superconducting gap, respectively.

STM topographic images were sample tilt corrected by subtracting a linear fit line by line. Two-dimensional FTs images were calculated using a standard Hamming window to reduce finite size effects with WsxM32 and Gwyddion analysis software.

ASSOCIATED CONTENT

**Supporting Information**. AFM characterization, Moiré pattern periodicity calculation, Superstructure modelling details, Simulations of the spatial dependence of the tunneling current on the heterostructures obtained when considering only the Bravais lattices, 12° MAH characterization, 12° MAH superconducting gap.

Corresponding Author

*E-mail: christoph.renner@unige.ch

ORCID

Jose Martinez-Castro: 0000-0001-7249-2567

Diego Mauro: 0000-0002-8290-1165

Árpád Pásztor: 0000-0001-8230-235X

Alessandro Scarfato: 0000-0002-5615-766X

Alberto F. Morpurgo: 0000-0003-0974-3620

Christoph Renner: 0000-0001-9882-681X

Acknowledgements

We thank G.Manfrini and A.Guipet for their technical assistance.

AFM and CR acknowledge financial support from the Swiss National Science Fundation through Div.II and Sinergia. AFM acknowledges support from the EU Graphene Flagship.

Notes

The authors declare no competing financial interest.

Author Contributions

AFM, CR, JMC, DM, AP designed the experiment. JMC carried out the scanning probe experiments, DM and IGL prepared the devices, and AP did the modelling. All authors did participate in the data analysis and manuscript writing. All authors have given approval to the final version of the manuscript. †These authors contributed equally.

A. M. WSXM: A Software for Scanning Probe Microscopy and a Tool for Nanotechnology. *Rev. Sci. Instrum.* **2007**, *78*, 13705.

TABLE OF CONTENTS GRAPHIC

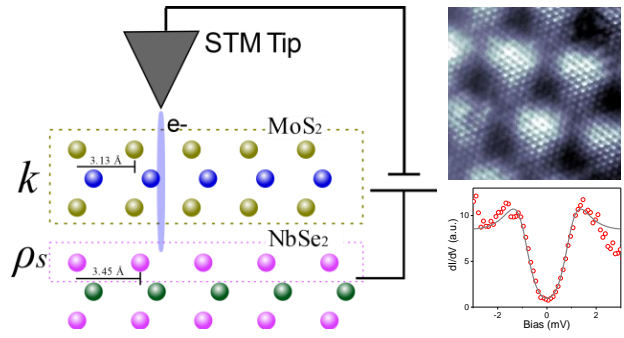